\documentclass[submission,copyright,creativecommons]{eptcs}
\providecommand{\event}{GANDALF 2019}
\usepackage{breakurl}             
\usepackage{underscore}           
\usepackage{amssymb}
\usepackage{xspace}
\usepackage{tikz}
\usepackage{amsthm}
\usepackage{bm}
\usepackage{multirow}

\newtheorem{theorem}{Theorem}
\newtheorem{proposition}[theorem]{Proposition}
\newtheorem{definition}[theorem]{Definition}

\newtheorem{example}[theorem]{Example}

\renewcommand{\iff}{\Leftrightarrow}

\newcommand{\mcomment}[1]{}

\newcommand{\A}{\ensuremath{\mathcal{A}}\xspace}

\newcommand{\dcsynth}{DCSynth\xspace}

\newcommand {\union} {\ensuremath{\cup}}

\renewcommand {\iff} {\ensuremath{\Leftrightarrow}}

\newcommand {\qddc} {QDDC\xspace}

\newcommand {\ang}[1] {\ensuremath{\langle#1\rangle}}
\newcommand {\sq}[1] {\ensuremath{[#1]}}
\newcommand {\dsq}[1] {\ensuremath{[[#1]]}}

\newcommand {\len}[1] {\ensuremath{len(#1)}}

\newcommand {\intv}[1] {\ensuremath{Intv(#1)}}
\newcommand {\slen} {\ensuremath{slen} }
\newcommand {\scount} {\ensuremath{scount} }
\newcommand {\sdur} {\ensuremath{sdur} }

\newcommand {\true} {\ensuremath{true}~}

\newcommand {\dom}[1] {\ensuremath{dom(#1)}}

\newcommand {\nat} {\ensuremath{\mathbb{N}}}

\newcommand{\oomit}[1]{}

\newcommand{\df}{=}

\newcommand{\DCSYNTH}{DCSynth\xspace}
\newcommand{\invariant}{\mbox{\bf AG\/}}
\newcommand{\MPNC}{MPS\xspace}
\newcommand{\GODSC}{MPHOS\xspace}

\begin{document}
\title{Specification and Optimal Reactive Synthesis of Run-time Enforcement Shields}
\author{Paritosh K. Pandya
\institute{Tata Institute of Fundamental Research \\
 Mumbai 400005, India}
\email{pandya@tifr.res.in}
\and
Amol Wakankar
\institute{Homi Bhabha National Institute}
\institute{Bhabha Atomic Research Centre \\
 Mumbai 400085, India}
\email{amolk@barc.gov.in}
}
\def\titlerunning{Specification and Optimal Reactive Synthesis of Run-time Enforcement Shields}
\def\authorrunning{P.K. Pandya and A. Wakankar}

\maketitle

\setcounter{footnote}{0}
\begin{abstract}
A system with sporadic errors (SSE) is a controller which produces high quality output but it may occasionally violate a critical requirement $REQ(I,O)$. A run-time enforcement shield is a controller which takes $(I,O)$ (coming from SSE) as its input, and it produces a corrected output  $O'$ which guarantees the invariance of requirement $REQ(I,O')$.  Moreover, the output sequence $O'$ must deviate from $O$ ``as little as possible'' to maintain the quality.  
In this paper, we give a method for logical specification of shields using formulas of logic {\emph Quantified Discrete Duration Calculus}(\qddc). The specification consists of a correctness requirement
$REQ$ as well as a hard deviation constraint $HDC$ which must both be mandatorily and invariantly satisfied by the shield. Moreover, we also use quantitative optimization to give a shield which minimizes the expected value of cumulative deviation in an $H$-optimal fashion. We show
how  tool \DCSYNTH implementing soft requirement guided synthesis can be used for automatic synthesis of shields from a given specification. Next, we give logical formulas specifying several notions of shields including the $k$-Stabilizing shield of Bloem \emph{et al.} \cite{BKKW15,KABHKTW17} as well as
the Burst-error shield of Wu \emph{et al.} \cite{WWZ16}, and a new $e,d$-shield. Shields can be automatically synthesized for all these specifications using the tool \DCSYNTH. We give experimental results showing the performance of our shield synthesis tool in relation to previous work. We also compare the performance of the shields synthesized under diverse hard deviation constraints in terms of their expected deviation and the worst case burst-deviation latency.
\end{abstract}

\section{Introduction}
\label{section:intro}

A system with sporadic errors (SSE) is a controller which produces high quality desirable output for any given input but it may sporadically violate a critical system requirement $REQ(I,O)$, where I and O are the set of input and output propositions. Many manually designed controllers have this character, as they embody designer's unspecified optimizations, however they may have obscure design errors. 
A run-time enforcement shield for a specified critical requirement $REQ(I,O)$ is a controller (Mealy machine) which receives both input and output $(I,O)$ generated by SSE. The shield produces a modified output $O'$ which is guaranteed to invariantly meet the critical requirement $REQ(I,O')$ (correct-by-construction). Moreover, in each run, the shield output $O'$ must deviate from the SSE output $O$ ``as little as possible'', to maintain the quality. This allows the shield to benefit from system designer's optimizations without having to formally specify these or to handle these in the synthesis. See Figure \ref{fig:safety-shield}.

A central issue in designing run-time enforcement shields is the underlying notion of ``deviating as little as possible'' from the SSE output. There 
are several different notions explored in the literature \cite{BKKW15,KABHKTW17,WWZ16,WWZ17}. 
In their pioneering paper, Bloem \emph{et al.} \cite{BKKW15} proposed the notion of  $k$-stabilizing shield which may deviate for at most $k$ cycles continuously under suitable assumptions. If assumptions are not met the shield may deviate arbitrarily. This was proposed as a hard requirement which must be mandatorily satisfied  by the shield in any behaviour. We call such constraints as \textbf{hard deviation constraints}. Konighofer \emph{et al} \cite{KABHKTW17} have proposed some variants of the $k$-stabilizing shield requirement with and without fail safe state, which are also hard deviation constraints. Specific shield synthesis algorithms have been developed for each of these constraints.

As our first main contribution, we propose a logical specification notation for \textbf{hard deviation constraints} using the formulas of an interval temporal logic \qddc. This logic allows us to succinctly and modularly specify regular properties \cite{Pan01a,MPW17,Pan01b}. With its counting constructs and interval based modalities, it can be used to conveniently specify both the correctness requirement $REQ(I,O)$ as well as the hard deviation constraint $HDC$.

Criticizing the inability of $k$-stabilizing shields in handling burst errors, Wu \emph{et al.} \cite{WWZ16, WWZ17} proposed a burst-error shield which enforces the invariance of
the correctness requirement, and it locally minimizes the measure of deviation between SSE output $O$ and the shield output $O'$, at each step.  An algorithm for the synthesis of such shields was given. We call such a shield as \emph{locally deviation minimizing}.

In this paper, as our second main contribution, we generalize the Wu technique to minimize the cumulative deviation more globally. 
An $H$-optimal shield which minimizes at each point the expected value of cumulative deviation in next $H$-steps of shield execution is computed. The cumulative deviation is averaged over all possible $H$ length inputs to arrive at the optimal estimate. A well known value iteration algorithm \cite{Bel57,Put94} for optimal policy synthesis of Markov Decision Processes allows us to compute such a shield.
We call such a shield as \emph{$H$-optimally deviation minimizing}. This is a powerful optimization and in the paper we  experimentally show its significant impact on performance of the shield. It may be noted that Wu's burst-error shield is obtained by selecting $H=0$. 

Finally, we propose a uniform method for synthesizing a run-time enforcement shield from given  logical specification $(REQ,HDC)$ and a horizon value (natural number) $H$. The resulting shield invariantly meets the correctness requirement $REQ$ as well as the hard deviation constraint $HDC$. Moreover, the shield is $H$-optimally deviation minimizing. The shield synthesis is carried out by using the soft requirement guided controller synthesis  tool \DCSYNTH \cite{WPM19}. This tool allows synthesis of $H$-optimal controllers from specified hard and soft \qddc\/ requirements. 

Using the proposed formalism, in the paper, we formulate several diverse notions of shields. These include a logical specification of 
Bloem's $k$-stabilizing shield and Wu's burst-error shield, as well as a new notion of $e,d$-shield. A uniform synthesis method using the tool \DCSYNTH
can be applied to obtain the corresponding run-time enforcement shields. It is notable that tool \DCSYNTH\/ uses an
efficient BDD-based semi-symbolic representation of automata/controllers with aggressive minimization. 
This allows the tool to scale better and to produce smaller sized shields. In the paper, we give an experimental evaluation of the 
performance of our \DCSYNTH tool and compare it with some previously reported studies in the literature.

With the ability to formulate shields with diverse hard deviation constraints,  it is natural to ask for a comparison of the {\bf performance} of these shields.
The performance must essentially measure the extent of deviation of the shield output from the SSE output. Towards this, we propose two 
measures of the shield performance.
\begin{itemize}
 \item  We compute the \emph{probability of deviation} in long run. For this, we assume that the input to the shield is fully random, with each input variable value chosen independently of the past and each other. While simplistic, this does provide some indication of the shield's effectiveness in average.
 \item We measure the \emph{worst case burst-deviation latency}. This gives the maximum number of consecutive deviations possible in the worst case. (If unbounded, we report $\infty$). A model checking technique implemented in a  tool CTLDC \cite{Pan05} allows us to compute this worst case latency. 
\end{itemize}
Tool \DCSYNTH\/ provides facilities for the computation of each of these performance measures for a synthesized shield. The reader may refer to the original papers on \DCSYNTH \cite{WPM19,PW19a} for details of techniques by which such performance can be measured. In this paper, we synthesize shields with different hard deviation constraints and we provide  a comparison of the performance of these shields. This allows us to draw some preliminary conclusions. Clearly, much wider experimentation is needed for firmer insight.

 The rest of the paper is organized as follows. Section \ref{section:qddc} describes the syntax and semantics of the logic \qddc. Section \ref{section:dcsynth-spec}  gives the syntax of \DCSYNTH specification and brief outline of the synthesis method. 
Section \ref{sec:shieldType} describes the various logical notions of shield specification. 
Section \ref{sec:performance} describes metrics to evaluate the shield performance and corresponding experimental results. 
In Section \ref{section:discussion}, we conclude the paper with discussion and related work.

\section{Preliminaries}
We provide a brief overview of logic \qddc\/ as well as the soft requirement guided $H$-optimal controller synthesis method implemented in tool \DCSYNTH.
This method and tool is applied to the problem of  run-time enforcement shield synthesis in this paper. 
The reader may refer to the original paper \cite{WPM19} for further details of these preliminaries.
\subsection{Quantified Discrete Duration Calculus (\qddc) Logic}
\label{section:qddc}
Let $PV$ be a finite non-empty set of propositional variables. 
Let $\sigma$ a non-empty finite word over the alphabet
$2^{PV}$. It has the form 
$\sigma=P_0\cdots P_n$ where $P_i\subseteq PV$ for each $i\in\{0,\ldots,n\}$. 
Let $\len{\sigma}=n+1$, 
$\dom{\sigma}=\{0,\ldots,n\}$, 
$\sigma[i,j]=P_i \cdots P_j$ and $\sigma[i]=P_i$.

The syntax of a \emph{propositional formula} over variables $PV$ is given by:
\[
\varphi := false\ |\ \true\ |\ p\in PV |\ !\varphi\ |\ \varphi~\&\&~\varphi\ |\ \varphi~||~\varphi
\]
with $\&\&, ||, !$ denoting conjunction, dis-junction and negation, respectively. Operators
such as $\Rightarrow$  and $\Leftrightarrow$ are defined as usual. 
Let $\Omega(PV)$ be the set of all propositional formulas over variables $PV$. 
%
Let $i\in\dom{\sigma}$. 
Then the satisfaction of propositional formula $\varphi$ at point $i$, denoted $\sigma,i\models\varphi$ is defined as usual and omitted here for brevity.

The syntax of a \qddc formula over variables $PV$ is given by: 
\[
\begin{array}{lc}
D:= &\ang{\varphi}\ |\ \sq{\varphi}\ |\ \dsq{\varphi}\ |\ 
D\ \verb|^|\ D\ |\ !D\ |\ D~||~D\ |\ D~\&\&~D\ \\ 
&ex~p.\ D\ |\ all~p.\ D\ |\ slen \bowtie c\ |\ scount\ \varphi \bowtie c\ |\ sdur\ \varphi \bowtie c 
\end{array} 
\]
where $\varphi\in\Omega(PV)$, $p\in PV$, 
$c ~ \in\nat$ and $\bowtie\in\{<,\leq,=,\geq,>\}$. 

An \emph{interval} over a word $\sigma$ is of the form $[b,e]$ 
where $b,e\in\dom{\sigma}$ and $b\leq e$. 
Let $\intv{\sigma}$ be the set of all intervals over $\sigma$.
Let $\sigma$ be a word over $2^{PV}$, let $[b,e]\in\intv{\sigma}$ be an interval. 
Then the satisfaction of a \qddc formula $D$  written as $\sigma,[b,e]\models D$, is defined inductively as follows:
\[
\begin{array}{lcl}
\sigma, [b,e]\models\ang{\varphi} & \mathrm{\ iff \ } & b=e \mbox{ and } \sigma,b\models \varphi,\\
\sigma, [b,e]\models\sq{\varphi} & \mathrm{\ iff \ } & b<e \mbox{ and }
                                           \forall b\leq i<e:\sigma,i\models \varphi,\\
\sigma, [b,e]\models\dsq{\varphi} & \mathrm{\ iff \ } & \forall b\leq i\leq e:\sigma,i\models \varphi,\\
\sigma, [b,e]\models D_1\verb|^| D_2 & \mathrm{\ iff \ } & \exists b\leq i\leq e:\sigma, [b,i]\models D_1\mbox{ and }\sigma,[i,e]\models D_2,\\
\end{array}
\]
with Boolean combinations $!D$, $D_1~||~D_2$ and $D_1~\&\&~D_2$  defined in the expected way. 
We call word $\sigma'$ a $p$-variant, $p\in PV$, of a word $\sigma$ 
if $\forall i\in\dom{\sigma},\forall q\neq p:q\in \sigma'[i]\iff q\in \sigma[i]$. 
Then $\sigma,[b,e]\models ex~p.~D \mathrm{\ iff \ } \sigma',[b,e]\models D$ for some 
$p$-variant $\sigma'$ of $\sigma$; and
$(all~p.~D) \Leftrightarrow (!ex~p.~!D)$. 

Entities \slen, \scount and \sdur are called \emph{terms}. 
The term \slen gives the length of the interval in which it is 
measured. Term $\scount\ \varphi$, where $\varphi\in\Omega(PV)$, counts 
the number of positions including the first and the last point 
in the interval under consideration where $\varphi$ holds. 
Formally, for $\varphi\in\Omega(PV)$ we have 
$\slen(\sigma, [b,e])=e-b$, and $\scount(\sigma,\varphi,[b,e])=\sum_{i=b}^{i=e}\left\{\begin{array}{ll}
					1,&\mbox{if }\sigma,i\models\varphi,\\
					0,&\mbox{otherwise.}
					\end{array}\right\}$.

We also define the following derived constructs: 
$pt=\langle true \rangle$, $ext=!pt$, $\bm{\langle \rangle D} = true\verb|^|D\verb|^|true$, $\bm{[]D}=(!\langle \rangle!D)$ and $\bm{pref(D)}=!((!D)\verb|^|true)$. 
Thus, $\sigma, [b,e] \models []D$ iff $\sigma, [b',e']\models D$ 
for all sub-intervals $b\leq b'\leq e'\leq e$ and $\sigma, [b,e] \models \mathit{pref}(D)$ 
iff $\sigma, [b,e']\models D$ for all prefix intervals $b \leq e' \leq e$.

Finally, we define
$\sigma,i \models D$ iff $\sigma, [0,i] \models D$, and 
$\sigma \models D$ iff $\sigma, [0,$ $\len{\sigma}-1] \models D$. Let $L(D) = \{ \sigma\mid\sigma \models D \}$, the set of behaviours accepted by $D$. Let $D$ be valid, denoted $\models_{dc} D$, iff $L(D) =
(2^{PV})^+$. Notice that $\sigma, i ~\models ~D$ denotes that the past of position $i$ satisfies the formula $D$.

\begin{example}
\label{exam:qddcExample} 
We give an example \qddc formula over propositions $\{p,q,r\}$ which specifies a typical recurrent reach-avoid behaviour required in many control systems. Intuitively, the formula $\varphi_{until}(n)$ holds at a position $i$ in the behaviour if, since the previous occurrence of $r$, the
proposition $p$ persists till an
occurrence of $q$. Moreover, $q$ must occur within $n$ time units from the last occurrence of $r$. For example, here $r$ may denote entering of enemy air-space, $p$ may denote that the UAV is invisible and $q$ may denote that the target is reached.
Let $\varphi_3$ abbreviate  $\varphi_{until}(3)$.
Figure \ref{fig:exampleUntil} gives a possible behaviour $\sigma$ where the last row gives the value of $\sigma,i \models  \varphi_3$ for each position $i$.
\begin{itemize}
\item $Until(p,q,n)$:\ \  
\verb|((slen<(n)) && [[p]]) | ~~\textbar\textbar$ \\
 \hspace*{4cm} $\verb|(((([p] | \textbar\textbar \verb| pt)^<q>) && slen<=n)^true)|. \\
The second disjunct holds for an interval $[b,e]$ provided $q$ occurs
at a position $b \leq j \leq e$ with $j \leq b+n$ and $p$ persists from
$b$ to $j-1$. E.g. in Figure \ref{fig:exampleUntil}, $\sigma,[5,9] \models until(p,q,3)$ with $j=8$. The
first disjunct holds for an interval $[b,e]$ provided $e-b < n$ and
$p$ holds throughout the interval. E.g. $\sigma,[11,12] \models until(p,q,3)$. Note that $\sigma,[2,4] \not \models until(p,q,3)$.

\item $SinceLast(p,D)$: \ \
\verb|!(true^(<p>^((slen=1^[[!p]])| \textbar\textbar \verb| pt) && !(D)))| \\
This formula fails to hold at position $i$ provided there is a previous (last)
occurrence of $p$ in the past of $i$, at say position $j \leq i$, and $D$ does not hold for the interval $[j,i]$.

\item Let $\varphi_{until}(n)$ be the \qddc formula \verb|SinceLast(r,(Until(p,q,n)))|.

Then, $\sigma,1 \models \varphi_3$ since there is no $r$ at any position
$j \leq 1$. Also, $\sigma, 9 \models  \varphi_3$ as, since the previous
occurrence of $r$ at position $5$, the proposition $p$ persists till $7$ and $q$ holds at $8$ (with $8 \leq 5+3$). Note also that $\sigma, 12 \models  \varphi_3$ since the previous $r$  occurs at $12$ (with $12 < 12+3$) and $\sigma,[12,12] \models \verb#[[ p ]]#$. Finally, $\sigma, 4 \not \models  \varphi_3$ as,
since the previous $r$ at position $4$, neither does $q$ occur in-between nor
do we have $\sigma,[2,4] \models \verb#[[ p ]]#$. \qed
\end{itemize}
\end{example}

\begin{figure}
\[
\begin{array}{|l|l|l|l|l|l|l|l|l|l|l|l|l|l|l|l|l|l|l|l|l|l|}
\hline
Position & 0 & 1 & 2 & 3 & 4 & 5 & 6 & 7 & 8 & 9 & 10 & 11 & 12 & 13 & 14 & 15 & 16 & 17 & 18 & 19 & 20 \\
\hline
r & 0 & 0 & 1 & 0 & 0 & 1 & 0 & 0 & 0 & 0 & 0 & 1 & 1 & 1 & 0& 0 &0&0&0&0&0\\
\hline
p & 0 & 0 & 0 &0 & 0 & 1 & 1 & 1 & 0 & 0 & 0 & 1 & 1 & 1 &1&1&1&1&1&1&1\\
\hline
q & 0 & 0 & 0 & 0 & 0 & 0 & 0 & 0 & 1 & 0 & 0 & 0 & 0 & 0 &0& 0 &0&0&0&0&0\\
\hline
\phi_{until}(3) & 1 & 1 & 0 &0 & 0 & 1 & 1 & 1 & 1 & 1 & 1 & 1 & 1 & 1 & 1 &1&0&0&0&0&0\\
\hline
\end{array}
\]
\caption{Example behaviour for $\phi_{until}(3)$}
\label{fig:exampleUntil}
\end{figure}
 
\begin{theorem}
\label{theorem:formula-automaton}
 \cite{Pan01a} For every formula $D$ over variables $PV$ we can construct a Deterministic Finite Automaton (DFA) $\A(D)$ over alphabet $2^{PV}$ 
such that $L(\A(D))=L(D)$. We call $\A(D)$ a \emph{formula automaton} for $D$ or the monitor automaton for $D$. \qed
\end{theorem}
A tool DCVALID implements this formula automaton construction in an efficient manner by internally using the tool MONA \cite{Mon02}. 
It gives {\em minimal, deterministic} automaton (DFA) for the formula $D$.
We omit the details here. However,  the reader may refer to several papers  on \qddc\/ for detailed description and examples of \qddc\/ specifications as well as  its model checking tool DCVALID \cite{Pan01a,MPW17,Pan01b}.

In the rest of the paper we consider \qddc\/  formulas and automata where variables $PV=I \cup O$ are partitioned into disjoint sets of input variables $I$ and output variables $O$. Such a formula/automaton specifies a relation between inputs and outputs.

For technical convenience, we define a notion of \textit{indicator variable} for a \qddc\/ formula
(regular property). The idea is that the indicator variable $w$ witnesses the truth of a formula $D$ at any point in execution. Thus, $Ind(D,w) \df pref(EP(w) \Leftrightarrow D)$. Here, $\mathbf{EP(w)} = (true \verb#^# \langle w \rangle)$, i.e. in a behaviour $\sigma$ and a position $i$, we have  $\sigma,i \models EP(w)$ iff $w \in \sigma[i]$. If $\sigma \models Ind(D,w)$ then for for any $i$, we have $\sigma,i \models D$ iff $w \in \sigma[i]$. Thus variable $w$ is true exactly at those positions where the past of the position satisfies $D$. These indicator variables can be used as auxiliary propositions in another formula using the notion of cascade composition $\ll$ defined below.
\begin{definition}[Cascade Composition]
\label{def:indDef}
Let $D_1, \ldots, D_k$ be \qddc\/ formulas over input-output variables $(I,O)$ and let $W=\{w_1, \ldots, w_k\}$ be the corresponding set of fresh indicator variables i.e. $(I \cup O) \cap W = \emptyset$. Let $D$ be a formula over variables $(I \cup  O \cup W)$. Then, the cascade composition $\ll$ and its equivalent \qddc\/ formula are as follows:
\[
  D \ll \langle Ind(D_1,w_1), \ldots, Ind(D_k,w_k) \rangle
 \quad  \df \quad D \land \bigwedge_{1 \leq i \leq k} ~ pref(EP(w_i) \Leftrightarrow D_i)
\]
This composition gives a formula over input-output variables $(I,O \cup W)$. \qed
\end{definition}
Cascade composition provides a useful ability to modularize a formula using auxiliary
propositions $W$ which witness other regular properties given as \qddc formulas. 

\begin{example}
\label{exam:cascade}
Consider a formula $\verb#D# = (\verb#scount dev <= 3#$) which holds at a point provided the proposition $\verb#dev#$ is $true$ at most 3 times in the entire past. Let formula $\verb#D1# = \verb#(true^<o#$ $\neq$ $\verb#o'>)#$ which holds at a point provided that the values of propositions \verb#o#  and \verb#o'# differ at that position. Then, $\verb#D# \ll \verb#Ind(D1,dev)#$ is equivalent to the formula $\verb#(scount dev <= 3) && #$  $\verb#pref(EP(dev) <=> D1)#$. This formula holds at a position \verb#i#, provided \verb#D1# holds at most $3$ time in the interval \verb#[0,i]#. That is \verb#o#$\neq$\verb#o'# for at-most $3$ positions in the interval \verb#[0,i]#. \qed
\end{example}

\subsection{Supervisors and Controllers}
Now we consider \qddc\/  formulas and automata where variables $PV=I \cup O$ are partitioned into disjoint sets of input variables $I$ and output variables $O$. We show how Mealy machines can be represented as special form of Deterministic finite automata (DFA). Supervisors and controllers are Mealy machines with special properties. This representation allows us to use the MONA DFA library \cite{Mon02} to efficiently compute supervisors and controllers in our tool \DCSYNTH.

\begin{definition}[Output-nondeterministic Mealy Machines]
\label{def:nondetmm}
A total and Deterministic Finite Automaton (DFA) over input-output alphabet $\Sigma=2^I \times 2^O$ is a tuple $A=(Q,\Sigma,s,\delta,F)$, as usual, with $\delta:Q \times 2^I \times 2^O \rightarrow Q$. An {\bf output-nondeterministic Mealy machine} is a DFA with a unique reject (or non-final) state $r$ 
which is a sink state i.e. $F= Q - \{r\}$ and $\delta(r,i,o)=r$ for all $i \in 2^I$, $o \in 2^O$. \qed
\end{definition}

Intuition is that the transitions from $q \in F$ to $r$ are
forbidden (and kept only for making the DFA total).
Language of any such Mealy machine is prefix-closed. 
Recall that for a Mealy machine, $F=Q-\{r\}$.
A Mealy machine is 
{\bf deterministic} if $\forall s \in F$, $\forall i \in 2^I$, $\exists$ at most one $o \in 2^O$  s.t. $\delta(s,i,o) \not=r$. 
An output-nondeterministic Mealy machine is called {\bf non-blocking} if $\forall s \in F$, $\forall i \in 2^I$ $\exists o \in 2^I$ s.t.
$\delta(s,i,o) \in F$. It follows that 
for all input sequences a non-blocking Mealy machine can produce one or more
output sequence without ever getting into the reject state. 

For a Mealy machine $M$ over variables $(I,O)$, its language $L(M) \subseteq (2^I \times 2^O)^*$. A word $\sigma \in L(M)$ can also be represented as pair $(ii,oo) \in ((2^I)^*,(2^O)^*) $ such that
$\sigma[k] = ii[k] \cup oo[k], \forall k \in dom(\sigma)$. Here $\sigma, ii, oo$ must have the same length. We will not distinguish between $\sigma$ and $(ii,oo)$ in the rest of the paper.
Also, for any input sequence $ii \in (2^I)^*$, we will define $M[ii] = \{ oo ~\mid~ (ii,oo) \in L(M) \}$. 

\begin{definition}[Controllers and Supervisors]
 An output-nondeterministic Mealy machine which is non-blocking is called a {\bf supervisor}.  A deterministic supervisor is called a {\bf controller}.  \qed
\end{definition}
The non-deterministic choice of outputs in a supervisor denotes unresolved decision.
The determinism ordering below allows supervisors to be refined into controllers.

\begin{definition}[Determinism Order and Sub-supervisor]
\label{def:detOrd}
 Given two supervisors $Sup_1, Sup_2$ we say that $Sup_2$ is {\em more deterministic} than $Sup_1$, denoted $Sup_1 \leq_{det} Sup_2$,
 iff $L(Sup_2) \subseteq L(Sup_1)$.  We call $Sup_2$ to be a {\bf sub-supervisor} of 
 $Sup_1$. \qed
 \end{definition}
Note that being supervisors, they are both non-blocking, and hence 
$\emptyset \subset Sup_2[ii] \subseteq Sup_1[ii]$ for any $ii \in (2^I)^*$.
The supervisor $Sup_2$ may make use of additional memory for resolving and pruning
the non-determinism in $Sup_1$. 

\subsection{\DCSYNTH Specification and  Controller Synthesis}
\label{section:dcsynth-spec}

This section gives a brief overview of the soft requirement guided controller synthesis method from \qddc formulas. 
The method is implemented in a tool \DCSYNTH. (See \cite{WPM19} for details). This method and the tool will be used for synthesis of run-time enforcement shields in the subsequent sections.
\begin{definition}
A supervisor  $Sup$ {\em realizes invariance} of  \qddc\/ formula $D$ over variables $(I,O)$, denoted as  ${\bf Sup~ \mathbf{realizes ~\invariant}~ (D)}$, provided $L(Sup) \subseteq L(D)$. Recall that, by the definition of supervisors, $Sup$ must be non-blocking.
The supervisor $Sup$ is called {\bf maximally permissive} provided for any supervisor
$Sup'$ such that ${\bf Sup'~ \mathbf{realizes ~\invariant}~ (D)}$, we have $Sup \leq_{det} Sup'$. Thus, no other supervisor with larger languages realizes the invariance of $D$.
This $Sup$ is unique up to language equivalence of automata, and the minimum state
maximally permissive supervisor is denoted by $\mathbf{\MPNC(D)}$. \qed
\label{def:realizableAndMPNC}
\end{definition}
A well-known greatest fixed point algorithm for safety synthesis over $\A(D)$ gives us 
$\MPNC(D)$ if it is realizable. We omit the details here (see \cite{WPM19}).
\begin{proposition}[\MPNC Monotonicity]
\label{prop:mpncSupervisor}
Given \qddc formulas $D_1$ and $D_2$ over variables $(I,O)$ such that $\models (D_1 \Rightarrow D_2)$, we have:
\begin{itemize}
\item $\MPNC(D_2)  \leq_{det} \MPNC(D_1)$, and
\item If $\MPNC(D_1)$ is realizable then $\MPNC(D_2)$ is also realizable.
\qed
\end{itemize}
\end{proposition}

A {\bf \DCSYNTH specification} is a tuple 
$(I,O,D^h, D^s)$ where $I$ and $O$ are the set of {\em input} and {\em output} variables, respectively. Formula $D^h$, called the \emph{hard requirement}, and formula $D^s$, called the \emph{soft requirement}, are \qddc\/ formulas over the set of propositions $PV=I \cup O$. 
The  objective in \DCSYNTH\/ is to synthesize a deterministic controller which (a) {\bf invariantly} satisfies the
hard requirement $D^h$, and (b) {\bf optimally} satisfies $D^s$ for as many inputs as possible.

The controller synthesis goes through following three stages.
\begin{enumerate}
 \item The \DCSYNTH specification $(I,O,D^h, D^s)$  is said to be realizable iff $\MPNC(D^h)$ is realizable (i.e. it exist). The synthesis method first computes the maximally permissive supervisor $\MPNC(D^h)$ realizing invariance of $D^h$. When clear from context we will abbreviate this as $\MPNC$.
 \item A sub-supervisor of $\MPNC(D^h)$  which satisfies $D^s$ for ``as many inputs as possible'' is computed.
This is formalized using a notion of {\bf $H$-optimality} w.r.t. the soft requirement $D^s$. We explain this only intuitively. The reader may refer to the original paper \cite{WPM19} for a formal definition of $H$-optimality and the synthesis algorithm. Let $H$ be a natural number called horizon. We construct the maximally permissive sub-supervisor  of $\MPNC(D^s)$, called $\GODSC(D^h,D^s,H)$, by pruning the non-deterministic choice of outputs in $\MPNC$ and retaining  only the outputs
which give \emph{the highest expected count of (intermittent) occurrence of $D^s$ over the next $H$ steps of execution}. 
This count is averaged  over all input sequences of length $H$. A well known value-iteration algorithm due to Bellman \cite{Bel57}, adapted from optimal
strategy synthesis for Markov Decision Processes \cite{Put94}, gives us the required $H$-optimal maximally permissive sub-supervisor. See the paper \cite{WPM19} for full details which are omitted here. Note that, by construction, $\MPNC(D^h) \leq_{det} \GODSC(D^h,D^s,H)$.
By Definition \ref{def:detOrd}, all the behaviours of $\GODSC$ will invariantly satisfy $D^h$ and the $\GODSC$ will be $H$-optimal with respect to $D^s$. 
When clear from context,
$\GODSC(D^h,D^s,H)$ will be abbreviated as $\GODSC$.
\item Both $\MPNC(D^h)$ and $\GODSC(D^h,D^s,H)$ are supervisors and they may be output-nondeterministic as there can be several optimal
outputs possible. Any controller obtained by arbitrarily resolving the output non-determinism in $\GODSC(D^h,D^s,H)$ will also be $H$-optimal.
In tool \DCSYNTH, we allow users to specify a preference ordering $Ord$ on the set of outputs  $2^O$. Any supervisor $Sup$ can be determinized by 
retaining only the highest ordered output among those permitted by $Sup$. This is denoted by $Det_{Ord}(Sup)$. In tool \DCSYNTH, the output ordering is specified by giving  a lexicographically ordered list of output variable literals, as illustrated in Example \ref{exam:lexoutput} below. 
This facility is used to determinize supervisors $\GODSC(D^h,D^s,H)$ and  $\MPNC(D^h)$ as required. These are denoted by  $Det_{ord}(\GODSC(D^h,D^s,H))$ and $Det_{Ord}(\MPNC(D^h))$.
\end{enumerate}

\begin{example}
\label{exam:lexoutput}
For a supervisor $Sup$ over variables $(I,\{p,q\})$, an output ordering can be
given as list ($!q > ~!p$),  Then, the determinization step will select the highest allowed  output  from the list ($p=false, q=false$), ($p=true, q=false$), ($p=false, q=true$), ($p=true, q=true$) in that order. This choice is made for each state and each input. \qed
\end{example}

In summary, given a \DCSYNTH\/ specification $(I,O,D^h, D^s)$, a horizon value $H$ and a preference ordering $ord$ on outputs $2^O$, the tool \DCSYNTH outputs
maximally permissive supervisors $\MPNC(D^h)$ and $\GODSC(D^h,D^s,H)$ as well as controllers $Det_{Ord}(\MPNC(D^h))$ and $Det_{ord}(\GODSC(D^h,D^s,H))$.

\paragraph{Extended \DCSYNTH specification:} \DCSYNTH supports the specification of soft requirements as an ordered list of formulas with user defined weights. This feature is used in the  synthesis of run-time enforcement shields.
The {\em extended \dcsynth specification} is a tuple 
$S=(I,O,D^h,\langle D^s_1:\theta_1,\cdots,D^s_k:\theta_k\rangle)$ 
where $I$ and $O$ are sets of {\em input} and {\em output} variables respectively. 
The \qddc formula $D^h$, which is over $I\union O$, specifies the {\em hard requirement} on the controller to be synthesized. 
The {\em soft requirement} $\langle D^s_1:\theta_1,\cdots, D^s_k:\theta_k \rangle$ 
is a list where each $D^s_i$ is a \qddc formula over $I\union O$.
$\theta_i\in\nat$ specifies the weight of the soft requirement $D^s_i$. The weight (reward) of a transition is sum of weights of each of the formula $D^s_i$
which holds on taking the transition.
The tool \DCSYNTH produces a supervisor, which maximizes the
cumulative expected value of this reward over next $H$-steps of execution. This cumulative reward is averaged over all input sequences of length $H$.

\section{Specification and Synthesis of Run-time Enforcement Shields}
\label{sec:shieldType}
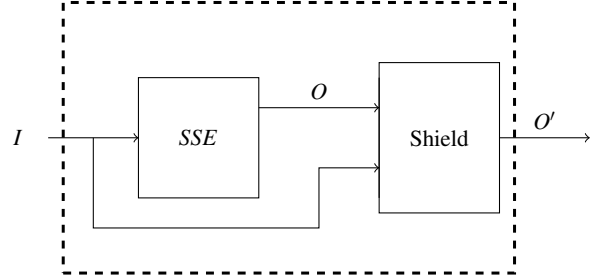
\begin{figure}[!h]
\begin{center}
\begin{tikzpicture}[scale=0.4, every node/.style={scale=0.8}]
\draw (2,0) node {$I$};
\draw[->] (3,0) -- (6,0);
\draw (6,-2) rectangle (10,2);
\draw (8,0) node {$SSE$};
\draw[->] (10,1) -- (14,1) node[midway,above] {$O$};
\draw[->] (4.5,0) -- (4.5,-3) -- (12,-3) -- (12,-1) -- (14,-1);
\draw (14,-2) -- (14, 2);
\draw (14,-2.5) rectangle (18,2.5);
\draw 
(16,0.0) node {Shield};
\draw[->] (18,0) -- (21,0) node[midway, above] {$O'$};
\draw[line width=.4mm, dashed] (3.5,-4.5) rectangle (18.5,4.5); 
\end{tikzpicture}
\end{center}
\caption{Run-time Enforcement Shield.}
\label{fig:safety-shield}
\end{figure}

Given a \textbf{correctness requirement} $REQ(I,O)$ as a \qddc\/ formula over input-output propositions $(I,O)$, a system with sporadic errors (SSE) may
fail to meet the requirement at some of the points in a behaviour $(ii,oo)$. (The reader may recall Definition \ref{def:nondetmm} and its following two paragraphs for the notation.)
A \textbf{run-time enforcement shield} is a Mealy machine with input 
variables $I \cup O$ and output variable $O'$. See Figure \ref{fig:safety-shield}.  For any input $(ii,oo)$ the shield produces a modified output $oo'$ such that $(ii,oo')$ invariantly satisfies the correctness requirement $REQ(I,O')$. Moreover, the output $oo'$ must deviate from the SSE output $oo$ as little as possible to maintain quality. 
There are several distinct notions of ``deviating as little as possible'' leading to different shields.

In this section, we give a logical framework  for specifying various shields by using the logic \qddc. We then provide 
an automatic synthesis of a run-time enforcement shield from its logical specification using the tool \DCSYNTH of the previous section. Thus, we achieve
a logical specification and a uniform synthesis method for shields.

Deviation constraints specify the extent of allowed deviation in a shield's behaviour. Our specification has  \textbf{hard deviation constraint} $HDC$ which must be mandatorily and invariantly satisfied  by the shield. (This is similar to the hard requirement in \DCSYNTH.)  We also define a 
canonical \textbf{soft deviation constraint} $Hamming(O,O')$ which will be useful in minimizing cumulative deviation during synthesis. 
Overall, a \textbf{shield specification} consists of a pair $(REQ,HDC)$. 

\subsection{Hard Deviation Constraints}
\label{sec:HDC}
Two indicator propositions, $SSEOK$ and $Deviation$ play an important role in formulating hard deviation constraints. Proposition $SSEOK$ indicates
whether the SSE is meeting the requirement $REQ(I,O)$ at the current position. Proposition $Deviation$ indicates whether at the current position, the shield output 
is different from the SSE output. Recall that in \DCSYNTH\/ specifications, the formula $Ind(D,w)$ defines a fresh output proposition $w$ which is true at a position provided the past of the position satisfies formula $D$ (see Definition \ref{def:indDef}). We use the following list of indicator definitions in formulating hard deviation constraints. Let, $O = \{o_1, \ldots, o_r\}$ and $O' = \{o_1', \ldots, o_r'\}$.
\[
INDDEF ~\df~ \left\langle  \begin{array}{l}
  Ind(~ REQ(I,O), ~SSEOK),  \\
  Ind(~ true \verb|^|\langle \lor_i (o_i \not= o'_i) \rangle, ~ Deviation)
\end{array}
\right\rangle
\]
A \textbf{hard deviation constraint} $HDC$ is a \qddc\/ formula over propositions $SSEOK$ and $Deviation$. It specifies a constraint on $Deviation$ conditional upon the behaviour of $SSEOK$.
In Subsection \ref{sec:shieldtype}, we will give a list of several different hard deviation constraints.

For shield synthesis using \DCSYNTH, we define the \qddc formula $HShield$ given in Equation \ref{eq:hardReqShield}) as the hard requirement over the input-output propositions $(I \cup O, O')$. Notice that in its formulation, we use the cascade composition from Definition \ref{def:indDef}. This allows us to modularize the specification into components  $REQ$ and $HDC$.
\begin{equation}
\label{eq:hardReqShield}
 HShield ~\df~ REQ(I,O') \land HDC(SSEOK,Deviation) \ll INDDEF
\end{equation}

The constraint (\qddc formula) $HShield$ must be invariantly satisfied by the shield. Tool \DCSYNTH gives us a maximally permissive supervisor
$\MPNC(HShield)$ with this property (See definition \ref{def:realizableAndMPNC}). This supervisor can be termed as \emph{shield-supervisor without deviation minimization} and it will be denoted by $\MPNC(REQ,HDC)$.

\subsection{Soft Deviation Constraint}
While $HDC$ already places some constraints on the permitted deviation, we can further optimize the deviation in 
supervisor $\MPNC(REQ,HDC)$ of the previous section. Quantitative optimization techniques from Markov Decision Processes can be used. (Stocasticity
comes from the distribution of inputs to the shield.) The tool \DCSYNTH allows us to specify such optimization 
using a list of soft requirement formulas with weights. The tool optimizes a supervisor to a sub-supervisor which maximizes the expected value of cumulative weight of soft requirements over next $H$-steps. 
This cumulative weight is averaged over all input sequences of length $H$. See Section \ref{section:dcsynth-spec} and \cite{WPM19} for further details. 

We make use of this $H$-optimal sub-supervisor computation to get a sub-supervisor  which minimizes the expected cumulative deviation over next $H$-steps. 
Given the set of output propositions $O= \{ o_1, \ldots, o_r\}$, consider the \DCSYNTH\/ soft-requirement
\begin{equation}
 Hamming(O,O') = \langle (true\verb|^|\langle o_1=o'_1 \rangle):1, ~ \ldots ~, (true\verb|^|\langle o_r=o'_r \rangle):1 \rangle
\end{equation}
Thus, non-deviation of any output variable $o_i=o'_i$ at current position contributes a reward $1$. This is summed over all output variables to give weight
(reward) of the soft requirement. Thus,
the weight of the soft requirement $Hamming(O,O')$ at any position $k$ in a word $(ii,oo,oo')$ is the value $(r - h)$ where $h$ is the  hamming distance between $oo[k]$ and $oo'[k]$. If $oo$ and $oo'$ perfectly match  at position $k$ then the weight at position $k$ is $r$, whereas if $oo$ and $oo'$ differ in values of say $p$ variables  at position $k$ then the weight at the position $k$ is $r-p$. 

By using $Hamming(O,O')$ as soft requirement and by selecting a horizon value $H$, we can apply the tool \DCSYNTH\/  to obtain a  sub-supervisor
\[
   \GODSC(~\MPNC(REQ,HDC),~Hamming(O,O'),~H~)
\]
of the supervisor $\MPNC(REQ,HDC)$. This sub-supervisor retains only the outputs which maximize the expected accumulated  weight of $Hamming(O,O')$ over next $H$ steps in future. 
This supervisor is called the \emph{shield-supervisor with deviation minimization} and denoted by $\GODSC(REQ,HDC,H)$.

\subsection{Determinization}
\label{sec:determinization}
The reader must note that both the shield-supervisors $\MPNC(REQ,HDC)$ and $\GODSC(REQ,HDC,H)$ are output non-deterministic. Multiple choice of outputs
may satisfy the hard deviation constraints while being $H$-optimal for the soft deviation constraint. 
Any arbitrary resolution of the output non-determinism will preserve the invariance guarantees and $H$-optimality (see \cite{WPM19}). 

In our method, we allow the user to specify a preference ordering $ord$ on the shield outputs $2^{O'}$. A lexicographically ordered list of output 
literals is given as explained in  Example \ref{exam:lexoutput}. A deterministic controller is obtained by  retaining only the highest ordered output from the non-deterministic choice of outputs offered by the supervisor. Thus,  given a preference ordering $ord$ we can obtain shields (\textbf{deterministic controllers})  $Det_{ord}(\MPNC(REQ,HDC))$ and $Det_{ord}(\GODSC(REQ,HDC,H))$.

In summary, given a correctness requirement $REQ(I,O)$ to be enforced by the shield, a hard deviation constraint $HDC(SSEOK,Deviation)$, a horizon value $H$ (for globally minimizing the deviation over next $H$ steps) and a preference ordering $ord$ on shield outputs $2^{O'}$, we can synthesize shields $Det_{ord}(\MPNC(REQ,HDC))$ and $Det_{ord}(\GODSC(REQ,HDC,H))$. When $ord,REQ,HDC,H$ are clear from context, these shields are referred to as $Shield\_NoDM$ (shield with no deviation minimization) and $Shield\_DM$ (shield with deviation minimization), respectively.

\subsection{Variety of Hard Deviation Constraints and Shield-Types}
\label{sec:shieldtype}
In Table \ref{tab:shieldTypes} below, we give a useful list of several different hard deviation constraints ($HDC$) as \qddc\/ formulas. These include the specifications of the burst-error shield of Wu \emph{et al.}
and  the $k$-stabilizing shield of Bloem \emph{et al. as} well as a new notion of $e,d$-shield. Labels $V_0$ to $V_3$ are used to identify these specifications in the experiments. Each of these $HDC$ can be used to synthesize  shields with or without deviation minimization as explained in the previous subsection.
\medskip

\begin{table}[h]
\caption{Variety of Hard Deviation Constraints}
\label{tab:shieldTypes}
\begin{center}
\begin{tabular}{|l|l|c|}
\hline
 & \textbf{ShieldType} & \textbf{HDC} \\ \hline
 $V_0$ & Burst-shield  & $true$  \\ \hline
 $V_1$ & $k$-shield  &  \verb|[]([[Deviation]]=>slen<k)| \\ \hline
 $V_2$ & $k$-stabilizing shield  &  \verb|[]([[SSEOK && Deviation]]=>slen<k)   &&| \\
                  & &  \verb|( []( (<!Deviation>^[[SSEOK]]) => [[!Deviation]] ) )|  \\ \hline

 $V_3$ & $e,d$-shield &  \verb|[]((scount !SSEOK <= e) => (scount Deviation <=d) &&| \\
                  & &  \verb|( []( (<!Deviation>^[[SSEOK]]) => [[!Deviation]] ) )| \\ \hline     
\end{tabular}
\end{center}           
\end{table}
We provide some explanation and comments on these specifications.
\begin{itemize}
 \item The proposition $SSEOK$ denotes that the $SSE$ is not making correctness error where as proposition $Deviation$ denotes that the shield is deviating from the $SSE$ output. The \qddc\/ formula \\
 \verb|( []( (<!Deviation>^[[SSEOK]]) => [[!Deviation]] ) )| states that in any observation interval, if the interval begins with no deviation, and there is no error by SSE during the interval, then there is no deviation throughout the interval. This property can be called $\mathit{NoSpuriousDeviation}$. It is included as a conjunct in $k$-Shield $V_2$ as well as $e,d$-Shield $V_3$. 
 \item Burst-shield ($V_0$) does not enforce any hard deviation constraint. Thus, only hard requirement  on the synthesized shield is to meet $REQ(I \cup O, O')$ invariantly. However, we can use this together with deviation minimization using
 the soft deviation constraint $Hamming(O,O')$. By taking horizon $H=0$, we obtain the burst sheild of Wu \emph{et al.} \cite{WWZ16} which locally optimizes deviation at 
 each step without any look-ahead into the future. Larger horizon values give superior shields which improve the probability of non-deviation in long run, as shown by our experiments which are reported later in this paper.
 \item A $k$-shield ($V_1$) specifies (as its hard deviation constraint) that for any observation interval the deviation can invariantly happen for at most $k$ cycles. Thus, a burst of deviation has length of at most $k$ cycles.
 The $k$-shield ($V_1$) specifies that this property must hold unconditionally. \emph{Such a specification is often unrealizable}. For example, if SSE makes consecutive errors for more than $k$ cycles, the shield may be forced to deviate for all of these cycles. Hence, several variants of the $V_1$ shield have been considered. 
 \item The $k$-stabilizing shield ($V_2$) specifies that the shield may deviate as long as $SSE$ makes errors (even burst errors). Once $SSE$ recovers from deviation 
 (indicated by $SSEOK$ becoming and remaining true), the shield may deviate for at most $k$ cycles. Thus, the shield must recover from
 deviation within $k$ cycles once $SSEOK$ is established and maintained. Also, there must be no spurious deviation due to conjunct $\mathit{NoSpuriousDeviation}$. This specification precisely gives the $k$-stabilizing shield without fail-safe state, originally defined by
 Konighofer \emph{et al.} \cite{KABHKTW17}. By a variation of this, the $k$-stabilizing shield with fail-safe state \cite{KABHKTW17} can also be specified but we omit this here.
 
\item We define a new notion of shield called $e,d$-shield ($V_3$). This states that in any observation interval if the count of errors by SSE (given by the term
\verb|(scount !SSEOK)|) is at most $e$ then the count of number of cycles with  deviations (given by the term
\verb|(scount Deviation)|) is at most $d$. Thus $e$ errors lead to at most $d$ deviations. Also, there is no spurious deviation due to the conjunct
$\mathit{NoSpuriousDeviation}$. 

\end{itemize}

It may be noted that irrespective of the shield type the synthesized shield have to meet the requirement $REQ(I,O')$ invariantly as specified by the formula $HShield$ (See Equation \ref{eq:hardReqShield}).

\begin{table}[!h]
\caption {Synthesis of Burst shield-$V_0$ with Deviation Minimization optimization using \DCSYNTH.  For each specification, the number of states of the resulting shield and time (in seconds) for synthesizing it are reported. 
For comparision, results for $k$-stabilizing shield synthesis and Burst-error shield synthesis are reproduced directly from Wu \emph{et al.} \cite{WWZ16}.
} 
\label{tab:dfaBasedComparision}
\smallskip

\centering
\scriptsize{
\begin{tabular}{|c||c|c|c|c||c|c|c|c|} 
\hline
\multirow{3}{*}{Specification}
&
\multicolumn{2}{l|}{$k$-Stabilizing shield}&\multicolumn{2}{l||}{Burst-error shield}&\multicolumn{4}{c|}{Burst shield $V_0$ with DM}\\\cline{2-3}\cline{4-5}\cline{6-9} 
&&&&&\multicolumn{2}{l|}{For H=0}&\multicolumn{2}{l|}{For H=10}\\
\cline{6-9}
&states&time&states&time&states&time&states&time\\\hline
\hline
Toyota Powertrain 
& 38        & 0.2       & 38 & 0.3 & 9 & 0.07 & 9 & 0.35\\\hline
Traffic light         
& 7        & 0.1       &7 & 0.2 & 4 & 0.008 & 4 & 0.059\\\hline
$F_{64}p$         
& 67   & 0.7       & 67 & 0.5 & 67 & 0.009 & 67 & 0.029 \\\hline
$F_{256}p$         
& 259    & 46.9 & 259 & 10.5 & 259 & 0.08 & 259 &  0.09\\\hline
$F_{512}p$         
& 515 & 509.1 & 515  & 54.4  & 515 & 0.24  & 515 & 0.26 \\\hline
G($\neg$ q) $\vee$  $F_{64}$(q $\wedge$ $F_{64}p$)    
& 67   & 0.8 & 67 & 0.6 & 67 & 0.015 & 67 & 0.06 \\\hline
G($\neg$ q) $\vee$  $F_{256}$(q $\wedge$ $F_{256}p$) 
& 259 & 46.2 & 259 & 10.7 & 259 & 0.16 & 259 & 0.27 \\\hline
G($\neg$ q) $\vee$  $F_{512}$(q $\wedge$ $F_{512}p$) 
& 515 & 571.7 & 515 & 54.5 & 515 & 0.77 & 515 & 0.91\\\hline
G(q $\wedge$ $\neg$ r $\rightarrow$ ($\neg$ r $\cup_{4}$ (p $\wedge$ $\neg$ r))) 
& 15 & 0.1 & 145 & 0.1 & 6 & 0.002 & 6 & 0.013\\\hline
G(q $\wedge$ $\neg$ r $\rightarrow$ ($\neg$ r $\cup_{8}$ (p $\wedge$ $\neg$ r))) 
& 109  & 0.2 & 5519 & 4.5 & 10 & 0.003 & 10 & 0.023 \\\hline
G(q $\wedge$ $\neg$ r $\rightarrow$ ($\neg$ r $\cup_{12}$ (p $\wedge$ $\neg$ r))) 
& 753 & 6.3 & 27338 & 1414.5 & 14 & 0.009 & 14 & 0.03 \\\hline \hline
AMBA G1+2+3        
& 22  & 0.1   & 22 & 0.1 & 7 & 0.002 & 7 & 0.01\\\hline
AMBA G1+2+4    
& 61 & 6.3 & 78 & 2.2 & 8 & 0.2 & 8 & 1.69 \\\hline
AMBA G1+3+4      
& 231  & 55.6 & 640 & 97.6 & 14 & 0.25  & 14 & 2.01 \\\hline
AMBA G1+2+3+5    
& 370 & 191.8  & 1405 & 61.8 & 12 & 0.017  & 13 & 0.105\\\hline
AMBA G1+2+4+5      
 & 101  & 3992.9 & 253 & 472.9 & 12 & 1.27  & 12 & 8.86 \\\hline
AMBA G4+5+6      
 & 252 & 117.9 & 205 & 26.4 & 18 & 0.86  & 18 & 7.99 \\\hline
AMBA G5+6+10     
& 329 & 9.8 & 396 & 31.4 & 27 & 3.7  & 27 & 36.14\\\hline
AMBA G5+6+9e4+10 
 & 455 & 17.6 & 804 & 42.1 & 46 & 5.58 & 46 & 52.96 \\\hline
AMBA G5+6+9e8+10 
 & 739 & 34.9 & 1349 & 86.8 & 64 & 7.44 & 64 & 70.73\\\hline
AMBA G5+6+9e16+10 
 & 1293 & 74.7 & 2420 & 189.7 & 100 & 11.3 & 100 & 105.2\\\hline
AMBA G5+6+9e64+10 
  & 4648 & 1080.8 & 9174 & 2182.5 & 316 & 37.17 & 316 & 202.52 \\\hline
AMBA G8+9e4+10   
 & 204 & 7.0 & 254 & 6.1 & 48 & 0.29 & 16 & 2.13\\\hline
AMBA G8+9e8+10   
 & 422 & 22.5 & 685 & 33.7 & 84 & 0.55  & 20 & 3.49 \\\hline 
AMBA G8+9e16+10   
  & 830 & 83.7 & 1736 & 103.1 & 156 & 1.02  & 28 & 6.32 \\\hline 
AMBA G8+9e64+10   
  & 3278 & 2274.2 & 7859 & 2271.5 & 588 & 5.96  & 76 & 24.89 \\\hline 
\end{tabular}
}
\end{table}

\section{Performance Measurement Metrics and Experiments}
\label{sec:performance}
In this section we give the experimental results for shield synthesis carried out in our framework. We first benchmark the performance of our tool and compare it with some other tools for shield synthesis in Section \ref{sec:shieldComparisonTool}. In Section \ref{sec:logicalShieldComparison} we define some performance measurement metrics for shields and we use these to compare various shield types.

\subsection{Performance of Tool \DCSYNTH\/ in Shield Synthesis }
\label{sec:shieldComparisonTool}
We have synthesized Burst-shield $V_0$ with deviation minimization using \DCSYNTH for all the benchmark examples given in \cite{WWZ16}.
The results are tabulated in Table~\ref{tab:dfaBasedComparision}. 
All our experiments were conducted on Linux (Ubuntu 18.04) system with Intel i5 64 bit, 2.5 GHz processor and 4 GB memory. 
The formula automata files of Wu \emph{et~al.}\cite{Wu16a} were used in place of \qddc\/ formulas for uniformity.
For a comparision with other tools, the results for the $k$-stabilizing shield synthesis and the Burst-error shield synthesis for the same examples are reproduced directly from Wu \emph{et al.} \cite{WWZ16}. As these are for unknown hardware setup, a direct comparison of the synthesis times with the \DCSYNTH synthesis times is only indicative.

As the table suggests,  in most of the cases, the shield synthesized by \DCSYNTH  compares favorably 
with the results reported in literature \cite{WWZ16}, 
both in terms of the size of the shield and the time taken for the synthesis.
Recall that \DCSYNTH uses aggressive minimization to obtain smaller shields. 
As an example, for the specification AMBA G5+6+9e64+10, our tool synthesizes a shield significantly faster and 
with smaller number of states than the existing tools\cite{BKKW15,WWZ16}.

\subsection{Comparison between various shield notions}
\label{sec:logicalShieldComparison}
For comparing the performance of shields synthesized with different shield types, we define the following performance metrics.

\paragraph{Expected Value of Non-deviation of a Shield in Long run:}
A shield is said to be in a non-deviating state if the shield output $O'$ matches the SSE output $O$. A proposition \verb|!Deviation| holds for such states.
We measure the probability of shield being in such states over its long runs, as described below.

Given a shield $S$ over input-output propositions $((I \cup O), O')$ and a \qddc\/ formula (regular property) $D$ over variables $I \cup O \cup O'$, we construct a {\em Discrete Time Markov Chain (DTMC)}, denoted as $M_{unif}(S,D)$, whose analysis allows us to measure the probability of $D$ holding in long runs (steady state) of $S$ under independent and identically distributed (iid) inputs. This value is called the expected value of $D$ holding in a shield $S$ and designated as $\mathbb{E}_{unif}(S,D)$. 

The construction of the desired DTMC 
is as follows. The product $S \times \A(D)$ gives a finite state automaton with the same behaviours as $S$. Moreover, it is in accepting state exactly when $D$ holds for the past behaviour. (Here $\A(D)$ works as a total deterministic monitor automaton for $D$ without restricting $S$). By assigning uniform discrete probabilities to all the inputs from any state, we obtain the DTMC $M_{unif}(S,D)$ along with a designated set of accepting states. The DTMC is in accepting state precisely when $D$ holds.
Standard techniques from Markov chain analysis allow us to compute the {\em probability} (Expected value) of being in the set of accepting states on long runs (steady state) of the DTMC. This gives us the desired value $\mathbb{E}_{unif}(S,D)$.
A leading probabilistic model checking tool MRMC implements this computation \cite{KZHHJ11}.
In \DCSYNTH, we provide a facility to compute $M_{unif}(S,D)$ in a format accepted by the tool MRMC. Hence, using \DCSYNTH\/ and  MRMC, we are able to compute  $\mathbb{E}_{unif}(S,D)$.

The expected value of a shield $S$ being in a non-deviating state over long runs can be computed as $\mathbb{E}_{unif} (S,$\verb|true^<!Deviation>|$)$. 


\paragraph{Worst Case Burst-Deviation Latency:}
The \emph{worst case burst-deviation latency} gives the maximum number of consecutive cycles for which the shield deviates even when the SSE is 
satisfying the requirement. Thus, it denotes the maximum length of an interval in the behaviour of the shield for which the formula ``$SSEOK ~\&\&~ Deviation$'' holds invariantly.

Given a Shield $S$ and a \qddc formula $D$, 
the latency goal $MAXLEN(D,S)$
computes
\[
 sup \{e-b ~~\mid~~ \rho,[b,e] \models D, ~\rho \in Exec(S) \}
\]
i.~e.~it computes the length of
the longest interval satisfying $D$ across all the executions of $S$. Thus, it computes the worst case span of behaviour fragments matching $D$ in $S$. 
Tool CTLDC \cite{Pan05} implements a model checking technique for computing MAXLEN(D,S). The worst case burst deviation latency of shield measures the maximum number of consecutive cycles having deviation in worst case.
The worst case burst-deviation latency of a shield $S$ can be computed as $MAXLEN([[SSEOK~ \&\& ~Deviation]],S)$.

\begin{table}[!h]
\caption{
Shield Synthesis for the formula $\varphi_{until}(5)$ of Example \ref{exam:qddcExample} with various shield types defined in Table \ref{tab:shieldTypes}
and their Performance comparison. The \emph{expected value of non-deviation in long run} and the \emph{worst case burst-deviation latency} are reported.
\oomit{State and time give the number of shield states and time taken (in seconds) by \DCSYNTH to compute the shield. Columns 
Experiments for Until Since  Example. The time for computation of each shield is given in seconds. 
The Expected value column shows the expected value of shield output not deviating from the system output i.e $\mathbb{E}_{unif}(S, ! Deviation)$. The Latency is computed for $MAXLEN[[SSEOK \&\& deviation]]$.}
}
\label{tab:expectedValAndLatency1}

\begin{center}
\begin{tabular}{|c|c||c|c||c|c|}
\hline
Sr. No. &	Shield Type &	States & Time &	Expected Value	& Latency \\
 \hline\hline
	\multicolumn{6}{|c|}{\textbf{Shield Synthesis of Requirement $\varphi_{until}(5)$ Without Deviation Minimization}}\\
	\hline
	1. & $V_0$\_NoDM & 18	& 0.004	& 0.25 & $\infty$ \\
\hline
	2. & $V_1$\_NoDM(k=1) &	Unrealizable &&& \\
	\hline		
	3. & $V_2$\_NoDM(k=1)&	14	& 0.004	& 0.7142793	& 1 \\
	\hline	
	4. & $V_1$\_NoDM(k=3)&	Unrealizable	&&& \\
	\hline		
	5. & $V_2$\_NoDM(k=3)	& 18 &	0.009	& 0.5982051	& 3\\
\hline
	6. & $V_3$\_NoDM(e=1,d=1)& 13 & 0.001 & 0.7499943 & 0\\
	\hline
	7. & $V_3$\_NoDM(e=1,d=2)& 26 & 0.005 & 0.7182475 & 1 \\
	\hline
	8. & $V_3$\_NoDM(e=1,d=3)& 40 & 0.008 & 0.6614611 & 2 \\
	\hline

	\multicolumn{6}{|c|}{\textbf{Shield Synthesis of Requirement $\varphi_{until}(5)$ With Deviation Minimization}}\\
	\hline
	9. & $V_1$\_DM(k=1) &	Unrealizable &&&\\
	\hline							
	10. & $V_0$\_DM(H=0)	& \multirow{6}{*}{13}	&	0.003 & \multirow{6}{*}{0.833252}	& \multirow{6}{*}{0}\\
	\cline{1-2}\cline{4-4}
	11. & $V_2$\_DM(k=1)(H=0)	&  &	0.005	&	& \\
	\cline{1-2}\cline{4-4}
	12. & $V_2$\_DM(k=3)(H=0) &		& 0.006	& & \\
	\cline{1-2}\cline{4-4}	
	13. & $V_3$\_DM(e=1,d=1)(H=0)& & 0.004 & & \\
	\cline{1-2}\cline{4-4}
	14. & $V_3$\_DM(e=1,d=2)(H=0)&  & 0.005 &   &  \\
	\cline{1-2}\cline{4-4}
	15. & $V_3$\_DM(e=1,d=3)(H=0)&  & 0.004 &  &   \\
	\hline

	16. & $V_0$\_DM(H=10)	& \multirow{6}{*}{8}	& 0.016	& \multirow{6}{*}{0.8571396}	& \multirow{6}{*}{0}\\
		\cline{1-2}\cline{4-4}
	17. & $V_2$\_DM(k=1)(H=10) &		& 0.01
	&  &	\\
	\cline{1-2}\cline{4-4}	
	18. & $V_2$\_DM(k=3)(H=10) &	 & 0.009
 &		& \\
 \cline{1-2}\cline{4-4}

	19. & $V_3$\_DM(e=1,d=1)(H=10)&  & 0.008 &  & \\
	\cline{1-2}\cline{4-4}
	20. & $V_3$\_DM(e=1,d=2)(H=10)&  & 0.012 &  & \\
	\cline{1-2}\cline{4-4}
	21. & $V_3$\_DM(e=1,d=3)(H=10)&  & 0.013 &  & \\
	
	\hline
\end{tabular}
\end{center}
\end{table}

\subsubsection{Experiments and Findings}
We can use the \emph{expected value of deviation} and the \emph{worst case burst-deviation latency}, defined above, for comparing the shields obtained using various shield-types defined in Section \ref{sec:shieldtype}. 
We synthesized  various shields for the correctness requirement $\varphi_{until}(n)$  given in Example \ref{exam:qddcExample} with $n=5$ and  the input-output
propositions $(\{r\},\{p,q\})$. The output propositions of synthesized shield are $\{p',q'\}$.
For each shield type $V_i$ given in Table \ref{tab:shieldTypes}, the deterministic shields  $V_i\_NoDM$ and $V_i\_DM$  were synthesized as outlined in the last paragraph of Section \ref{sec:determinization}. Here  $V_i\_NoDM$ denotes shield synthesized without deviation minimization where as $V_i\_DM$ denotes
the shield obtained with deviation minimization optimization. 
The shield-supervisors were determinized with the preference ordering ($!q'>!p'$) on outputs.

Table \ref{tab:expectedValAndLatency1} gives the results obtained. We report the number of states of the shield along with the time taken (in seconds) by the tool \DCSYNTH\/ to compute the shield. Moreover, for comparing the performance of the resulting shields, their \textbf{Expected Value of non-deviation} as well as the \textbf{worst case burst-deviation latency} are reported in the table under the columns titled Expected Value and Latency, respectively.

It is observed that  with deviation minimization optimization, several different shield types resulted in identical shields, although the time to synthesize them differed.  For example, shields in rows numbered 10 to 15 are identical.
We indicate such a situation by merging the corresponding rows to a single cell. We give our findings below. 
\begin{itemize}
\item The $k$-shield ($V_1$) is unrealizable as expected. See its description in Section \ref{sec:shieldtype} for an explanation. All the other shield types
are found to be realizable.

\item For shield synthesis without deviation minimization, we obtain distinct shields with distinct performance for each shield type. The Burst shield
($V_0$) has the poorest performance (expected non-deviation $0.25$ and latency $\infty$) as it enforces trivial hard deviation requirement $true$.
The best performance is obtained for the newly defined $e,d$-shield type $V_3$ by choosing $e=d$. This gives $0.74$ as the expected value of non-deviation and worst case latency of $0$ cycles. With increased difference $d-e$ the performance degrades. Similarly in $k$-stabilizing shield ($V_2$) the performance degrades with increase in the value  of $k$, as expected.

\item The performance of the shield considerably improves with the deviation minimization (DM) optimization. Expected value of 0.85 compares
well against the best value of 0.74 without deviation minimization. Also burst-deviation latency drops to $0$ with DM. 
We also notice that the performance improves with increase in the horizon value when using DM. This is intuitively clear as the tool performs global 
optimization across larger number of steps of look-ahead with increased horizon.

\item For shield synthesis with deviation minimization optimization, all the different shield types $V_0, V_2, V_3$ resulted in identical shield for 
a given value of horizon $H$. Thus shields in rows 10-15 (synthesized with $H=0$) and rows 16-21 (synthesized with $H=10$)  are found to be identical.
This shows that deviation minimization effectively supersedes the different hard deviation guarantees provided by the $HDC$. While this is not theoretically
guaranteed, our experience with robust controller synthesis also indicates the overwhelming effectiveness of the DM-like optimization \cite{PW19a}.
\end{itemize}

\section{Discussion and Related Work}
\label{section:discussion}

In this paper we have presented a logical framework for specifying error-correcting run-time enforcement shields using formulas of logic \qddc. The specification contains a correctness requirement $REQ$, specifying the desired input-output relation to be maintained,  as well as a hard deviation constraint $HDC$ which specifies a constraint on deviation between the system output and the shield output. Our shield synthesis gives a shield which invariantly satisfies both $REQ$ and $HDC$. Moreover, a powerful optimization globally minimizes the cumulative deviation between the system and the shield output.

The idea of error-correcting run-time enforcement shield was proposed in the pioneering work of Bloem \emph{et al.} \cite{BKKW15}, where the notion of $k$-stabilizing shield (with a synthesis algorithm) was proposed. This was further enhanced by Konighofer \emph{et al.} \cite{KABHKTW17}.
Extension of shield synthesis to liveness properties has also been explored in this paper.
 Wu \emph{et al.} \cite{WWZ16,WWZ17} defined the burst shield which is capable of handling burst errors. Moreover, they proposed optimizing the shield with the choice of output which locally minimizes the deviation at each stage. In this paper, we have enhanced this with global optimization of cumulative deviation across next $H$ steps.

In our method, the shield is logically specified using \qddc\/ formulas and a uniform method for the synthesis of the shield is proposed. A tool \DCSYNTH\/ implements the synthesis method. Logic \qddc\/  \cite{Pan01a,Pan01b,MPW17} with its interval logic modalities, threshold counting constraints, regular expression like constructs
and second-order quantification over temporal variables provides a very rich vocabulary to specify both the system requirements and the deviation constraints.
Logic \qddc\/ is a discrete time version of Duration Calculus proposed by Zhou, Hoare and Ravn \cite{CHR91,ZH04} with known automata theoretic decision and model checking procedures \cite{Pan01a,CP03,SPC05,SP05}. Using the proposed technique, we have specified the $k$-stabilizing shield of Konighofer \emph{et al.} \cite{KABHKTW17}, the burst shield of Wu \emph{et al.} \cite{WWZ16,WWZ17}, as well as a new $e,d$-shield. Moreover, we have measured the performance of the
shields resulting from these different criteria in terms of the expected value of deviation in long runs, as well as the worst case burst deviation latency. Our experiments show an overwhelming impact of global deviation minimization on the quality of the shield.
At the same time, hard deviation constraints provide a conditional hard guarantee on the worst case deviation. Hence, the combination of hard deviation constraint together with global minimization of deviation is useful.

Konighofer \emph{et al.} \cite{KABHKTW17} as well as Ehlers and Topku \cite{ET14} propose controller/shield synthesis technique for optimal achievable value of parameter $k$ in a regular specification. By contrast, our current method requires $k$ to be specified. In our future work, we will address similar optimal parametric synthesis from parameterized \qddc\/ specifications. 
\bibliographystyle{eptcs}
\bibliography{awRef1}

\begin{thebibliography}{10}
\providecommand{\bibitemdeclare}[2]{}
\providecommand{\surnamestart}{}
\providecommand{\surnameend}{}
\providecommand{\urlprefix}{Available at }
\providecommand{\url}[1]{\texttt{#1}}
\providecommand{\href}[2]{\texttt{#2}}
\providecommand{\urlalt}[2]{\href{#1}{#2}}
\providecommand{\doi}[1]{doi:\urlalt{http://dx.doi.org/#1}{#1}}
\providecommand{\bibinfo}[2]{#2}

\bibitemdeclare{book}{Bel57}
\bibitem{Bel57}
\bibinfo{author}{R.~E. \surnamestart Bellman\surnameend}
  (\bibinfo{year}{1957}): \emph{\bibinfo{title}{Dynamic Programming}}.
\newblock \bibinfo{publisher}{Princeton Univ. Press}.

\bibitemdeclare{inproceedings}{BKKW15}
\bibitem{BKKW15}
\bibinfo{author}{Roderick \surnamestart Bloem\surnameend},
  \bibinfo{author}{Bettina \surnamestart K{\"o}nighofer\surnameend},
  \bibinfo{author}{Robert \surnamestart K{\"o}nighofer\surnameend} \&
  \bibinfo{author}{Chao \surnamestart Wang\surnameend} (\bibinfo{year}{2015}):
  \emph{\bibinfo{title}{Shield Synthesis: - Runtime Enforcement for Reactive
  Systems}}.
\newblock In \bibinfo{editor}{Christel \surnamestart Baier\surnameend}, editor:
  {\sl \bibinfo{booktitle}{TACAS}}, {\sl \bibinfo{series}{LNCS}}
  \bibinfo{volume}{9035}, \bibinfo{publisher}{Springer}, pp.
  \bibinfo{pages}{533--548}, \doi{10.1007/978-3-662-46681-0_51}.

\bibitemdeclare{inproceedings}{CP03}
\bibitem{CP03}
\bibinfo{author}{Gaurav \surnamestart Chakravorty\surnameend} \&
  \bibinfo{author}{Paritosh~K. \surnamestart Pandya\surnameend}
  (\bibinfo{year}{2003}): \emph{\bibinfo{title}{Digitizing Interval Duration
  Logic}}.
\newblock In \bibinfo{editor}{Warren~A. \surnamestart Hunt\surnameend} \&
  \bibinfo{editor}{Fabio \surnamestart Somenzi\surnameend}, editors: {\sl
  \bibinfo{booktitle}{{CAV}}}, {\sl \bibinfo{series}{LNCS}}
  \bibinfo{volume}{2725}, \bibinfo{publisher}{Springer}, pp.
  \bibinfo{pages}{167--179}, \doi{10.1007/978-3-540-45069-6\_17}.

\bibitemdeclare{book}{ZH04}
\bibitem{ZH04}
\bibinfo{author}{Zhou \surnamestart Chaochen\surnameend} \&
  \bibinfo{author}{Michael~R. \surnamestart Hansen\surnameend}
  (\bibinfo{year}{2004}): \emph{\bibinfo{title}{Duration Calculus - {A} Formal
  Approach to Real-Time Systems}}.
\newblock \bibinfo{series}{Monographs in Theoretical Computer Science. An
  {EATCS} Series}, \bibinfo{publisher}{Springer},
  \doi{10.1007/978-3-662-06784-0}.

\bibitemdeclare{article}{CHR91}
\bibitem{CHR91}
\bibinfo{author}{Zhou \surnamestart Chaochen\surnameend},
  \bibinfo{author}{C.~A.~R. \surnamestart Hoare\surnameend} \&
  \bibinfo{author}{A.~P. \surnamestart Ravn\surnameend} (\bibinfo{year}{1991}):
  \emph{\bibinfo{title}{A Calculus of Durations}}.
\newblock {\sl \bibinfo{journal}{Inf. Process. Lett.}}
  \bibinfo{volume}{40}(\bibinfo{number}{5}), pp. \bibinfo{pages}{269--276},
  \doi{10.1016/0020-0190(91)90122-X}.

\bibitemdeclare{inproceedings}{ET14}
\bibitem{ET14}
\bibinfo{author}{R\"{u}diger \surnamestart Ehlers\surnameend} \&
  \bibinfo{author}{Ufuk \surnamestart Topcu\surnameend} (\bibinfo{year}{2014}):
  \emph{\bibinfo{title}{Resilience to Intermittent Assumption Violations in
  Reactive Synthesis}}.
\newblock In: {\sl \bibinfo{booktitle}{HSCC}}, \bibinfo{series}{HSCC '14},
  \bibinfo{publisher}{ACM}, \bibinfo{address}{New York, NY, USA}, pp.
  \bibinfo{pages}{203--212}, \doi{10.1145/2562059.2562128}.

\bibitemdeclare{article}{KZHHJ11}
\bibitem{KZHHJ11}
\bibinfo{author}{J.~\surnamestart Katoen\surnameend}, \bibinfo{author}{I.~S.
  \surnamestart Zapreev\surnameend}, \bibinfo{author}{E.~M. \surnamestart
  Hahn\surnameend}, \bibinfo{author}{H.~\surnamestart Hermanns\surnameend} \&
  \bibinfo{author}{D.~N. \surnamestart Jansen\surnameend}
  (\bibinfo{year}{2011}): \emph{\bibinfo{title}{The ins and outs of the
  probabilistic model checker MRMC}}.
\newblock {\sl \bibinfo{journal}{Performance Evaluation}} \bibinfo{volume}{68},
  pp. \bibinfo{pages}{89--220}, \doi{10.1016/j.peva.2010.04.001}.
\newblock
  \urlprefix\url{http://www.sciencedirect.com/science/article/pii/S0166531610000660}.

\bibitemdeclare{article}{Mon02}
\bibitem{Mon02}
\bibinfo{author}{N.~\surnamestart Klarlund\surnameend},
  \bibinfo{author}{A.~\surnamestart M\o{}ller\surnameend} \&
  \bibinfo{author}{M.~I. \surnamestart Schwartzbach\surnameend}
  (\bibinfo{year}{2001}): \emph{\bibinfo{title}{{MONA} Implementation Secrets}}
  \bibinfo{volume}{2088}, pp. \bibinfo{pages}{182--194}.
\newblock \doi{10.1007/3-540-44674-5_15}.

\bibitemdeclare{article}{KABHKTW17}
\bibitem{KABHKTW17}
\bibinfo{author}{Bettina \surnamestart K{\"o}nighofer\surnameend},
  \bibinfo{author}{Mohammed \surnamestart Alshiekh\surnameend},
  \bibinfo{author}{Roderick \surnamestart Bloem\surnameend},
  \bibinfo{author}{Laura \surnamestart Humphrey\surnameend},
  \bibinfo{author}{Robert \surnamestart K{\"o}nighofer\surnameend},
  \bibinfo{author}{Ufuk \surnamestart Topcu\surnameend} \&
  \bibinfo{author}{Chao \surnamestart Wang\surnameend} (\bibinfo{year}{2017}):
  \emph{\bibinfo{title}{Shield synthesis}}.
\newblock {\sl \bibinfo{journal}{FMSD}}
  \bibinfo{volume}{51}(\bibinfo{number}{2}), pp. \bibinfo{pages}{332--361},
  \doi{10.1007/s10703-017-0276-9}.

\bibitemdeclare{inproceedings}{SP05}
\bibitem{SP05}
\bibinfo{author}{Shankara~Narayanan \surnamestart Krishna\surnameend} \&
  \bibinfo{author}{Paritosh~K. \surnamestart Pandya\surnameend}
  (\bibinfo{year}{2005}): \emph{\bibinfo{title}{Modal Strength Reduction in
  Quantified Discrete Duration Calculus}}.
\newblock In: {\sl \bibinfo{booktitle}{{FSTTCS}}}, {\sl \bibinfo{series}{LNCS}}
  \bibinfo{volume}{3821}, \bibinfo{publisher}{Springer}, pp.
  \bibinfo{pages}{444--456}, \doi{10.1007/11590156\_36}.

\bibitemdeclare{inproceedings}{MPW17}
\bibitem{MPW17}
\bibinfo{author}{Raj~Mohan \surnamestart Matteplackel\surnameend},
  \bibinfo{author}{Paritosh~K. \surnamestart Pandya\surnameend} \&
  \bibinfo{author}{Amol \surnamestart Wakankar\surnameend}
  (\bibinfo{year}{2017}): \emph{\bibinfo{title}{Formalizing Timing Diagram
  Requirements in Discrete Duration Calculus}}.
\newblock In: {\sl \bibinfo{booktitle}{SEFM 2017}}, {\sl
  \bibinfo{series}{LNCS}} \bibinfo{volume}{10469}, \bibinfo{publisher}{Springer
  International Publishing}, pp. \bibinfo{pages}{253--268},
  \doi{10.1007/978-3-319-66197-1_16}.

\bibitemdeclare{inproceedings}{Pan01b}
\bibitem{Pan01b}
\bibinfo{author}{Paritosh~K. \surnamestart Pandya\surnameend}
  (\bibinfo{year}{2001}): \emph{\bibinfo{title}{Model Checking {CTL}*[{DC}]}}.
\newblock In: {\sl \bibinfo{booktitle}{TACAS}}, {\sl \bibinfo{series}{LNCS}}
  \bibinfo{volume}{2031}, \bibinfo{publisher}{Springer}, pp.
  \bibinfo{pages}{559--573}, \doi{10.1007/3-540-45319-9_38}.

\bibitemdeclare{inproceedings}{Pan01a}
\bibitem{Pan01a}
\bibinfo{author}{Paritosh~K. \surnamestart Pandya\surnameend}
  (\bibinfo{year}{2001}): \emph{\bibinfo{title}{Specifying and deciding
  quantified discrete-time duration calculus formulae using DCVALID}}.
\newblock In: {\sl \bibinfo{booktitle}{RTTOOLS (affiliated with CONCUR 2001)}},
  \bibinfo{publisher}{CiteSeer}.

\bibitemdeclare{article}{Pan05}
\bibitem{Pan05}
\bibinfo{author}{Paritosh~K. \surnamestart Pandya\surnameend}
  (\bibinfo{year}{2005}): \emph{\bibinfo{title}{Finding Extremal Models of
  Discrete Duration Calculus formulae using Symbolic Search}}.
\newblock {\sl \bibinfo{journal}{Electronic Notes in Theoretical Computer
  Science}} \bibinfo{volume}{128}(\bibinfo{number}{6}), pp. \bibinfo{pages}{247
  -- 262}, \doi{10.1016/j.entcs.2005.04.015}.
\newblock
  \urlprefix\url{http://www.sciencedirect.com/science/article/pii/S1571066105002471}.
\newblock \bibinfo{note}{AVoCS 2004}.

\bibitemdeclare{article}{PW19a}
\bibitem{PW19a}
\bibinfo{author}{Paritosh~K. \surnamestart Pandya\surnameend} \&
  \bibinfo{author}{Amol \surnamestart Wakankar\surnameend}
  (\bibinfo{year}{2019}): \emph{\bibinfo{title}{Specification and Reactive
  Synthesis of Robust Controllers}}.
\newblock {\sl \bibinfo{journal}{CoRR}} \bibinfo{volume}{abs/1905.11157}.
\newblock \urlprefix\url{http://arxiv.org/abs/1905.11157}.

\bibitemdeclare{book}{Put94}
\bibitem{Put94}
\bibinfo{author}{Martin~L. \surnamestart Puterman\surnameend}
  (\bibinfo{year}{1994}): \emph{\bibinfo{title}{Markov Decision Processes:
  Discrete Stochastic Dynamic Programming}}, \bibinfo{edition}{1st} edition.
\newblock \bibinfo{publisher}{John Wiley \& Sons, Inc.}, \bibinfo{address}{New
  York, NY, USA}, \doi{10.1002/9780470316887}.

\bibitemdeclare{inproceedings}{SPC05}
\bibitem{SPC05}
\bibinfo{author}{Babita \surnamestart Sharma\surnameend},
  \bibinfo{author}{Paritosh~K. \surnamestart Pandya\surnameend} \&
  \bibinfo{author}{Supratik \surnamestart Chakraborty\surnameend}
  (\bibinfo{year}{2005}): \emph{\bibinfo{title}{Bounded Validity Checking of
  Interval Duration Logic}}.
\newblock In: {\sl \bibinfo{booktitle}{{TACAS}}}, {\sl \bibinfo{series}{LNCS}}
  \bibinfo{volume}{3440}, \bibinfo{publisher}{Springer}, pp.
  \bibinfo{pages}{301--316}, \doi{10.1007/978-3-540-31980-1\_20}.

\bibitemdeclare{article}{WPM19}
\bibitem{WPM19}
\bibinfo{author}{Amol \surnamestart Wakankar\surnameend},
  \bibinfo{author}{Paritosh~K. \surnamestart Pandya\surnameend} \&
  \bibinfo{author}{Raj~Mohan \surnamestart Matteplackel\surnameend}
  (\bibinfo{year}{2019}): \emph{\bibinfo{title}{{DCSYNTH:} {A} Tool for Guided
  Reactive Synthesis with Soft Requirements, (To Appear in Proc. VSTTE 2019)}}.
\newblock {\sl \bibinfo{journal}{CoRR}} \bibinfo{volume}{abs/1903.03991}.
\newblock \urlprefix\url{http://arxiv.org/abs/1903.03991}.

\bibitemdeclare{article}{Wu16a}
\bibitem{Wu16a}
\bibinfo{author}{Meng \surnamestart Wu\surnameend} (\bibinfo{year}{2016}):
  \emph{\bibinfo{title}{iShield2 Synthesizer}}.
\newblock {\sl
  \bibinfo{journal}{https://bitbucket.org/mengwu/shield-synthesis/}}.
\newblock \urlprefix\url{https://bitbucket.org/mengwu/shield-synthesis/}.

\bibitemdeclare{inproceedings}{WWZ17}
\bibitem{WWZ17}
\bibinfo{author}{Meng \surnamestart {Wu}\surnameend},
  \bibinfo{author}{H.~\surnamestart {Zeng}\surnameend},
  \bibinfo{author}{C.~\surnamestart {Wang}\surnameend} \&
  \bibinfo{author}{H.~\surnamestart {Yu}\surnameend} (\bibinfo{year}{2017}):
  \emph{\bibinfo{title}{INVITED: Safety guard: Runtime enforcement for
  safety-critical cyber-physical systems}}.
\newblock In: {\sl \bibinfo{booktitle}{DAC}}, \bibinfo{publisher}{ACM}, pp.
  \bibinfo{pages}{1--6}, \doi{10.1145/3061639.3072957}.

\bibitemdeclare{inproceedings}{WWZ16}
\bibitem{WWZ16}
\bibinfo{author}{Meng \surnamestart Wu\surnameend}, \bibinfo{author}{Haibo
  \surnamestart Zeng\surnameend} \& \bibinfo{author}{Chao \surnamestart
  Wang\surnameend} (\bibinfo{year}{2016}): \emph{\bibinfo{title}{Synthesizing
  Runtime Enforcer of Safety Properties Under Burst Error}}.
\newblock In \bibinfo{editor}{Sanjai \surnamestart Rayadurgam\surnameend} \&
  \bibinfo{editor}{Oksana \surnamestart Tkachuk\surnameend}, editors: {\sl
  \bibinfo{booktitle}{{NFM}}}, {\sl \bibinfo{series}{LNCS}}
  \bibinfo{volume}{9690}, \bibinfo{publisher}{Springer}, pp.
  \bibinfo{pages}{65--81}, \doi{10.1007/978-3-319-40648-0_6}.

\end{thebibliography}
\end{document}